# Hierarchical Equations of Motion Solved with the Multiconfigurational Ehrenfest Ansatz

Zhecun Shi, Huiqiang Zhou, Lei Huang, Rixin Xie, and Linjun Wang*

Zhejiang Key Laboratory of Excited-State Energy Conversion and Energy Storage,

Department of Chemistry, Zhejiang University, Hangzhou 310058, China

*Email: ljwang@zju.edu.cn

**Abstract:** Being a numerically exact method for the simulation of dynamics in open quantum systems, the hierarchical equations of motion (HEOM) still suffers from the curse of dimensionality. In this study, we propose a novel MCE-HEOM method, which introduces the multiconfigurational Ehrenfest (MCE) ansatz to the second quantization formalism of HEOM. Here, the MCE equations of motion are derived from the time-dependent variational principle in a composed Hilbert-Liouville space, and each MCE coherent-state basis can be regarded as having an infinite hierarchical tier, such that the truncation tier of auxiliary density operators in MCE-HEOM can also be considered to be infinite. As demonstrated in a series of representative spin-boson models, our MCE-HEOM significantly reduces the number of variational parameters and could efficiently handle the strong non-Markovian effect, which is difficult for conventional HEOM due to the requirement of a very deep truncation tier. Compared with MCE, MCE-HEOM reduces the number of effective bath modes and circumvents the initial samplings for finite temperature, eventually resulting in a huge reduction of computational cost.

## I. INTRODUCTION

The concept of open quantum systems adopts the philosophy of dividing a total system under investigation into two parts, i.e., system and bath. Namely, the system part often contains fewer degrees of freedom (DoFs), while the effect of macroscopic environment on the system can be depicted by the bath with almost an infinite number of DoFs. Such system-bath models have been widely used in physics and chemistry,[1–5] such as nuclear magnetic resonance,[6,7] high-energy physics,[8–10] light-harvesting complexes[11–16] and nonlinear spectroscopy.[17–22] In some cases, they are also denoted as impurity models, as the system can be considered as an impurity with respect to the bulk background.[23]

Albeit the importance of open quantum systems has been well recognized, how to build a universal and efficient theory has been in debate for a long time. An *ab initio* solution to this problem is starting from the total system-bath perspective and may trace off the bath DoFs to get the equations of motion (EOMs) for the reduced density matrix (RDM) of the system. However, it is impossible to obtain closed EOMs for the RDM in general cases. The system-bath partition also induces a nonlocal interaction on the time axis with respect to the RDM, which is known as the non-Markovian effect.[24] Being treated as an impurity model, some well-established methods can be used, such as the numerical renormalization group (NRG)[25–27] and continuous time-quantum Monte Carlo (CT-QMC).[28–31] After considering proper approximations, probably the most widely used formalism is the perturbative quantum master equations (QME).[32–37] As a perturbative method, however, the most widely used 2nd-order QME often breaks down when the system-bath coupling is too strong such that the truncation is not valid.

When the bath is consisted of a nearly infinite number of DoFs, its statistics can be generally depicted by the Gaussian distribution based on the law of large numbers, which corresponds to the modeling of bath as harmonic oscillators with linear system-bath couplings. Based on the analytical properties of harmonic oscillators, the bath DoFs can be integrated out analytically in the path integral form. This gives the famous Feynman-Vernon influence functional,[38,39] which contains a nonlocal memory kernel form in time and prevents analytical evaluation. Based on the truncation properties of the memory kernel, Makri and coworkers developed the quasi-adiabatic path integral (QUAPI) approach[40–46] to evaluate the RDM by iterative tensor multiplication. In particular, by decomposing the bath correlation function (BCF) as a linear combination of exponential functions, the path integral form of the RDM can be differentiated and results in the hierarchical equations of motion (HEOM, or hierarchical quantum master equations, HQME) approach.[47–54] The non-Markovian effect and system-bath coupling are considered by the auxiliary density operators (ADOs), which possess a hierarchical structure and can be regarded as Fock-space vectors in the occupation number representation for the effective dissipative modes defined by the exponential expansion of the BCF.[55] Since the proposal of HEOM, it has been widely used in many open quantum system problems[56–73] and is probably the most powerful method to deal with bilinear Hamiltonians. As the RDM is on the top tier of the hierarchy, its time evolution is influenced by the ADOs in a complex structure and the physical meaning of ADOs should also be properly clarified. Zhu and coworkers have derived a recursive relation for the ADOs and related them to the moments of the solvent modes.[74] Yan introduced a novel quasi-particle "dissipaton" to explain the decomposition of the BCF and reinterpreted the recursive relation of ADOs as the representation of the dissipaton algebra.[75–77] In addition, HEOM can be derived based on the generalized Wick's theorem and the generalized diffusion equation. Thereby, this formalism is called dissipaton equations of motion (DEOM), while the ADOs are also renamed as the dissipaton density operators (DDOs). Based on DEOM, correlation functions of the solvent modes and solvation momentums can be calculated easily in the DDO space.[78,79]

Although HEOM has established a powerful framework for open quantum system dynamics, the curse of dimensionality inherent in the quantum mechanism still exists due to the number scaling of ADOs.[53,55] It has already been indicated by Tanimura that HEOM can be mapped into a Schrödinger-like equation with an effective Hamiltonian (or an effective Liouvillian super-operator in the Liouville space), as long as the ADOs are properly aligned as the expansion coefficients of a state vector.[47] As HEOM can be regarded as a novel effective time-dependent Schrödinger equation (TDSE) or Liouville equation, low-rank approximations and tensor decomposition approaches[80–88] can be used to solve it. The first work combining HEOM with tensor decomposition was carried out by Shi and coworkers.[89] They employed the matrix product state (MPS) to represent the ADO tensor, which could reduce the exponential computational cost of HEOM. Similar framework has also been proposed by Borrelli using the twin space form for density matrix.[90] Note that different tensor network structures have been applied to the HEOM methodology when studying open quantum system problems,[91,92] such as charge transport,[93–96] molecule scattering near metal surface,[97] and polariton chemistry.[98–100]

Due to the rapid development of machine learning and quantum computing, new approaches can also be considered to treat the curse of dimensionality. Recently, the neural network architecture has been used to unravel the entanglement between correlated DoFs in open quantum systems since the first attempt of applying the restricted Boltzmann machine to study the dynamics of spin chains.[101–106] As a direct extension, the combination of machine learning and HEOM has also been investigated and the multi-dimensional ADO tensors can be represented by the restricted Boltzmann machine ansatz.[107,108] In parallel, quantum computing has also attracted many attentions in the theoretical chemistry community.[109–112] Designing proper quantum circuits to realize the evolution of HEOM is also a rational inspiration. Despite the non-Hermitian nature of the effective Hamiltonian in HEOM hampers the use of unitary quantum gates, many strategies have been proposed to overcome this problem, such as the linear combination of unitaries (LCU),[113,114] Sz.-Nagy unitary dilation[115,116] and the time-dependent variational principle (TDVP).[117–120] Li and Lv proposed a quantum algorithm based on LCU to realize the non-Hermitian evolution in HEOM.[121,122] Dan and coworkers have implemented HEOM with the dilation method to simulate open quantum systems on quantum computers.[123] For a review of recently developed machine learning and quantum computing methods for open quantum systems, we recommend ref. 108.

Quantum dynamics methods based on coherent states (or Gaussian wave packets) have also been widely used in theoretical chemistry, such as the multiple spawning,[124–127] the multiconfigurational Ehrenfest (MCE) dynamics[128–135] and the multiple Davydov ansatz.[136–145] The semiclassical feature of coherent states indicates a trajectory-based interpretation for quantum dynamics, which seems to be the most natural description of bath modes. Moreover, coherent states are the eigenstates of the annihilation operators and can be used in conjecture with the second quantized many-particle Hamiltonians. Based on the properties of coherent states, path integral-based analytical approaches for condensed-phase Hamiltonians can be derived.[146,147] As a variational ansatz, the tangent space of coherent states also supports a Kähler manifold with complex structure such that TDVP can be directly applied.[148–155] Thereby, the coherent states (as well as the linear combination of coherent states) can be compared with tensor network states and regarded as the CANDECOMP/PARAFAC (CP) decomposition of multidimensional tensors[156] or as a type of large-site MPS form for bosonic modes.[157] All these facts indicate that the coherent states might have similar capabilities as tensor network states.

In this paper, we propose the MCE-HEOM method, which adopts the MCE ansatz in conjecture with HEOM. We investigate whether MCE can significantly reduce the exponential computational cost as tensor network state methods. Based on the effective Hamiltonian given by HEOM, the EOMs of MCE can be similarly derived as for the traditional TDSE. As a result, routine MCE procedures can be applied directly in MCE-HEOM. Compared with tensor network state approaches, the MCE ansatz contracts all the effective bath modes into a large MPS site such that we do not need to worry about the orbital ordering and the structure of the tensor network state.[91,92] Moreover, operators expressed in the normal order can be directly evaluated in the framework of coherent states such that constructions of the optimal matrix product operator (MPO) or tensor product operator (TPO) can also be circumvented,[158,159] which significantly simplifies the implementation. In particular, the coherent state basis is highly advantageous for low-frequency modes with more significant semiclassical features, which corresponds to the challenging strong non-Markovian effect in HEOM.[92,160] Compared to MCE dynamics to solve the traditional TDSE, the number of effective bath modes is markedly reduced by the HEOM formalism, which results in simplified system-bath entanglement. Moreover, the initial sampling[128] or the thermal field treatment[139,161] for finite temperature is not required in MCE-HEOM. Although the Liouville space dynamics is utilized, the computational cost is generally smaller when the system part is not too large.

In Sec. II, we describe the HEOM formalism and the effective Hamiltonian given by properly aligning the ADOs as the coefficients of a state vector. The EOMs and related computational procedures of MCE-HEOM are also given. Numerical results of the spin-boson models in different parameter regions are presented in Sec. III. Finally, we make the conclusions and perspectives for future development in Sec. IV.

**II. THEORY**

**A. The second quantization form of HEOM**

In general, the total Hamiltonian of an open quantum system can be expressed as

$$\hat{H}_\mathrm{T} = \hat{H}_\mathrm{S} + \hat{H}_\mathrm{B} + \hat{H}_\mathrm{SB}, \tag{1}$$

where $\hat{H}_\mathrm{S}$ and $\hat{H}_\mathrm{B}$ are respectively the system and bath parts, and $\hat{H}_\mathrm{SB}$ stands for the system-bath interaction. Within the HEOM formalism, $\hat{H}_\mathrm{S}$ can take arbitrary forms while $\hat{H}_\mathrm{B}$ is assumed to be described by a collection of harmonic oscillators,

$$\hat{H}_\mathrm{B} = \sum_j \omega_j \left( \hat{a}_j^\dagger \hat{a}_j + \frac{1}{2} \right), \tag{2}$$

where the creation and annihilation operators for these bosonic modes satisfy the canonical commutation relation $\left[\hat{a}_j, \hat{a}_k^\dagger\right] = \delta_{jk}$. $\hat{H}_\mathrm{SB}$ is taken to be

$$\hat{H}_\mathrm{SB} = \hat{Q} \otimes \hat{F}. \tag{3}$$

Here, the dissipative system mode operator $\hat{Q}$ can also be arbitrarily set, while the corresponding bath operator $\hat{F}$ is assumed to be linear,

$$\hat{F} = \sum_j g_j \left( \hat{a}_j^\dagger + \hat{a}_j \right). \tag{4}$$

As $\hat{F}$ is a linear combination of bath modes and couples directly to the system, it is also named as the solvent mode operator. Although we here only consider the simplest single-mode coupling in Eq. (3), the approaches below can be generalized directly to multi-mode coupling cases with $\hat{H}_\mathrm{SB} = \sum_\alpha \hat{Q}_\alpha \otimes \hat{F}_\alpha$. In Eq. (4), the system-bath coupling coefficients $\{g_j\}$ can be defined through the spectral density as

$$J(\omega) = \pi \sum_j g_j^2 \delta(\omega - \omega_j). \tag{5}$$

For open quantum system studies, the spectral density is often taken to be a continuous function in order to obtain dissipative dynamics. The fluctuation-dissipation theorem relates the BCF to the spectral density as[162]

$$C(t) = \left\langle \hat{F}_\mathrm{B}(t) \hat{F}_\mathrm{B}(0) \right\rangle_\mathrm{B} = \frac{1}{\pi} \int_{-\infty}^{\infty} d\omega\, e^{-i\omega t} \frac{J(\omega)}{1 - e^{-\beta\omega}}, \tag{6}$$

where $\hat{F}_B(t) = e^{i\hat{H}_B t}\hat{F}e^{-i\hat{H}_B t}$, $\langle \cdot \rangle_B = \text{Tr}_B\left[\cdot \hat{\rho}_B^{eq}\right] = \text{Tr}_B\left[\cdot e^{-\beta \hat{H}_B}/Z_B\right]$, $Z_B = \text{Tr}_B\left[e^{-\beta \hat{H}_B}\right]$, and $\beta = 1/k_B T$. When the initial state can be factorized as

$$\hat{\rho}_T(0) = \hat{\rho}_S(0) \otimes \hat{\rho}_B^{eq}, \tag{7}$$

the time evolution of the system RDM, $\hat{\rho}_S(t) = \text{Tr}_B[\hat{\rho}_T(t)]$, can be analytically derived by path integral techniques[38,39] or the Wick's theorem for time-ordered operators[163]

$$\hat{\rho}_S(t) = e^{-i\hat{H}_S t}\mathcal{T}\left[\hat{\mathcal{F}}(t,0)\hat{\rho}_S(0)\right]e^{i\hat{H}_S t}, \tag{8}$$

where $\mathcal{T}$ is the time-ordering operator and $\hat{\mathcal{F}}(t,0)$ is the influence functional[55]

$$\hat{\mathcal{F}}(t,0) = \exp\left[-\int_0^t d\tau \int_0^\tau d\tau' \left(\hat{Q}_S^>(\tau) - \hat{Q}_S^<(\tau)\right)\left(C(\tau-\tau')\hat{Q}_S^>(\tau') - C^*(\tau-\tau')\hat{Q}_S^<(\tau')\right)\right], \tag{9}$$

where $\hat{O}^>\hat{A} = \hat{O}\hat{A}$, $\hat{O}^<\hat{A} = \hat{A}\hat{O}$ and $\hat{Q}_S(t) = e^{-i\hat{H}_S t}\hat{Q}e^{i\hat{H}_S t}$. To obtain a set of closed differential equations, the BCF is generally decomposed by exponential functions as

$$C(t) = \sum_{k=1}^{K} d_k e^{-\gamma_k t}, \tag{10}$$

where $K$ is the number of exponents and $\{d_k\}$ ($\{\gamma_k\}$) are the expansion coefficients (the exponents). The corresponding time-reversal formula is written as[48,49,51,53]

$$C(-t) = C^*(t) = \sum_{k=1}^{K} d_{\bar{k}}^* e^{-\gamma_k t}, \tag{11}$$

with $\bar{k}$ being defined by $\gamma_{\bar{k}} \equiv \gamma_k^*$ as $\{\gamma_k\}$ are either real or complex conjugated paired. The ADOs can then be defined as

$$\hat{\rho}_\mathbf{n}^{(n)}(t) = e^{-i\hat{H}_S t}\mathcal{T}\left[\prod_{k=1}^{K}\left(n_k!|d_k|^{n_k}\right)^{-1/2}\left(i\int_0^t d\tau e^{-\gamma_k(t-\tau)}\left(d_k\hat{Q}_S^>(\tau) - d_{\bar{k}}^*\hat{Q}_S^<(\tau)\right)\right)^{n_k}\hat{\mathcal{F}}(t,0)\hat{\rho}_S(0)\right]e^{i\hat{H}_S t}, \tag{12}$$

where $\mathbf{n} = (n_1, \cdots, n_K)$ and $n = \sum_{k=1}^{K} n_k$. Note that Eq. (12) has used the scaling scheme proposed by Shi and coworkers,[164,165] which highly reduces the number of ADOs used in HEOM and is also important for the derivation of the second quantization form below. By design, the system RDM is the zeroth-order ADO $\rho_\mathbf{0}^{(0)}$ at the top of the hierarchy tier. Based on Eq. (12), HEOM can be derived by differentiating $\hat{\rho}_{\mathbf{n}}^{(n)}$ with respect to time,

$$\dot{\hat{\rho}}_{\mathbf{n}}^{(n)} = -\mathrm{i}\left[\hat{H}_\mathrm{S}, \hat{\rho}_{\mathbf{n}}^{(n)}\right] - \sum_{k=1}^{K} n_k \gamma_k \hat{\rho}_{\mathbf{n}}^{(n)} - \mathrm{i}\sum_{k=1}^{K}\sqrt{(n_k+1)|d_k|}\left[\hat{Q}, \hat{\rho}_{\mathbf{n}_k^+}^{(n+1)}\right] \\ -\mathrm{i}\sum_{k=1}^{K}\sqrt{n_k/|d_k|}\left(d_k \hat{Q}\hat{\rho}_{\mathbf{n}_k^-}^{(n-1)} - d_k^* \hat{\rho}_{\mathbf{n}_k^-}^{(n-1)}\hat{Q}\right), \tag{13}$$

where $\mathbf{n}_k^+$ and $\mathbf{n}_k^-$ denote $(n_1,\cdots,n_k+1,\cdots,n_K)$ and $(n_1,\cdots,n_k-1,\cdots,n_K)$, respectively. For a specific truncation tier $L$, the total number of ADOs is[164]

$$N(L,K) = \sum_{n=0}^{L} \frac{(n+K)!}{n!K!}, \tag{14}$$

which implies an almost prohibitive scaling with respect to $L$ and $K$ for complex cases.

The subscript $\mathbf{n}$ resembles the occupation configuration of bosonic modes, which indicates a second quantization form of HEOM. To further simplify the formalism of HEOM, we may follow the method proposed by Tanimura[50] and introduce an auxiliary Fock state $|\mathbf{n}\rangle = |n_1\rangle \otimes |n_2\rangle \otimes \cdots \otimes |n_K\rangle$ for the effective bath modes. We also consider the bosonic creation-annihilation operators with

$$\hat{b}_k^\dagger |\mathbf{n}\rangle = \sqrt{n_k+1}\,|\mathbf{n}_k^+\rangle, \tag{15}$$

$$\hat{b}_k |\mathbf{n}\rangle = \sqrt{n_k}\,|\mathbf{n}_k^-\rangle, \tag{16}$$

$$\hat{n}_k |\mathbf{n}\rangle \equiv \hat{b}_k^\dagger \hat{b}_k |\mathbf{n}\rangle = n_k |\mathbf{n}\rangle, \tag{17}$$

such that a state vector which encodes the information of ADOs and the hierarchical structure can be accordingly defined as

$$|\Psi(t)\rangle = \sum_{\mathbf{n}} \hat{\rho}_{\mathbf{n}}^{(n)}(t) \otimes |\mathbf{n}\rangle. \tag{18}$$

Taking the time derivative of $|\Psi(t)\rangle$ and combining Eqs. (13) and (15)-(17) give[90]

$$|\dot{\Psi}(t)\rangle = -\left(i\mathcal{L}_S + \sum_{k=1}^{K}\gamma_k \hat{n}_k\right)|\Psi(t)\rangle - i\sum_{k=1}^{K}\sqrt{|d_k|}\hat{b}_k\left(\hat{Q}^> - \hat{Q}^<\right)|\Psi(t)\rangle$$

$$-i\sum_{k=1}^{K}\frac{1}{\sqrt{|d_k|}}\hat{b}_k^\dagger\left(d_k\hat{Q}^> - d_{\bar{k}}^*\hat{Q}^<\right)|\Psi(t)\rangle \quad (19)$$

$$\equiv -i\hat{H}_{\text{eff}}|\Psi(t)\rangle.$$

where $\mathcal{L}_S \equiv \hat{H}_S^> - \hat{H}_S^<$ and the effective Hamiltonian is defined as

$$\hat{H}_{\text{eff}} = \mathcal{L}_S - i\sum_{k=1}^{K}\gamma_k \hat{n}_k + \sum_{k=1}^{K}\sqrt{|d_k|}\hat{b}_k\left(\hat{Q}^> - \hat{Q}^<\right) + \sum_{k=1}^{K}\frac{1}{\sqrt{|d_k|}}\hat{b}_k^\dagger\left(d_k\hat{Q}^> - d_{\bar{k}}^*\hat{Q}^<\right). \quad (20)$$

Although Eq. (19) is equivalent to the original form of HEOM in Eq. (13), the second quantization form indicates that HEOM actually transforms the original bath modes into a new set of effective modes with the dissipative feature. The state vector $|\Psi(t)\rangle$ is now in the space $\mathcal{H}_S \otimes \mathcal{H}_S^* \otimes \mathcal{H}_D$,[122] where $\mathcal{H}_S \otimes \mathcal{H}_S^*$ corresponds to the reduced system part and $\mathcal{H}_D$ corresponds to the auxiliary state vector $|\mathbf{n}\rangle$ for the effective modes. Thereby, Eq. (19) implies that the effective modes in the Hilbert space interact with the system DoFs in the Liouville space. From a numerical perspective, however, Eq. (19) is still a linear equation and can be solved by those methods used for the TDSE or the Liouville equation (e.g., tensor network state approaches and MCE).

**B. HEOM with the MCE ansatz**

Eq. (19) reveals that the new effective bath modes provided by HEOM are also bosonic. Thereby, we may introduce the coherent state $|\mathbf{z}\rangle = |z_1\rangle \otimes |z_2\rangle \otimes \cdots \otimes |z_K\rangle$ for the creation-annihilation operators of the effective bath modes

$$\hat{b}_k|\mathbf{z}\rangle = z^{(k)}|\mathbf{z}\rangle, \quad (21)$$

where the normalized coherent state for each effective mode can be given by the corresponding occupation number states $\{|n\rangle\}$ as

$$|z\rangle = e^{-\frac{|z|^2}{2}} \sum_{n=0}^{\infty} \frac{z^n}{\sqrt{n!}} |n\rangle. \qquad (22)$$

Based on the coherent state basis, the state vector can also be written by the MCE ansatz[128]

$$|\Psi(t)\rangle = \sum_{\mu=1}^{N_b} \sum_{ij} A_{ij}^{\mu}(t) |\mathbf{z}_{\mu}(t)\rangle \otimes |ij\rangle\rangle \equiv \sum_{\mu} |\psi_{\mu}(t)\rangle, \qquad (23)$$

where $|ij\rangle\rangle \equiv |i\rangle\langle j|$ corresponds to the system density matrix in the Liouville space,[17] and $N_b$ is the total number of coherent states used to expand the state vector (i.e., the rank of the CP decomposition). $|\psi_{\mu}(t)\rangle$ is the component of the $\mu$-th coherent state,

$$|\psi_{\mu}(t)\rangle = |\mathbf{z}_{\mu}(t)\rangle \sum_{ij} A_{ij}^{\mu}(t) |ij\rangle\rangle, \qquad (24)$$

$$|\mathbf{z}_{\mu}(t)\rangle = |z_{\mu}^{(1)}(t)\rangle \otimes |z_{\mu}^{(2)}(t)\rangle \otimes \cdots \otimes |z_{\mu}^{(K)}(t)\rangle, \qquad (25)$$

In this paper, the MCEv1 ansatz is adopted, and the EOMs of the vibronic amplitudes $\{A_{ij}^{\mu}\}$ are given by the TDVP in the target space $\mathcal{H}_S \otimes \mathcal{H}_S^* \otimes \mathcal{H}_D$,[128,166]

$$|\dot{\Psi}_T\rangle = -i\hat{\mathcal{P}}_T \left( \hat{H}_{\text{eff}} |\Psi\rangle - i|\dot{\Psi}_z\rangle \right). \qquad (26)$$

Here, the time derivative in the tangent space $|\dot{\Psi}_T\rangle$, the tangent-space projector $\hat{\mathcal{P}}_T$ and the time derivative provided by the coherent states $|\dot{\Psi}_z\rangle$ are respectively given by

$$|\dot{\Psi}_T\rangle = \sum_{\mu=1}^{N_b} \sum_{ij} \dot{A}_{ij}^{\mu}(t) |\mathbf{z}_{\mu}(t)\rangle \otimes |ij\rangle\rangle, \qquad (27)$$

$$\hat{\mathcal{P}}_T = \sum_{\mu\nu} |z_{\mu}\rangle \Omega^{\mu\nu} \langle z_{\nu}| \otimes \sum_{ij} |ij\rangle\rangle\langle\langle ij|, \qquad (28)$$

$$|\dot{\Psi}_z\rangle = \sum_{\mu=1}^{N_b} \sum_{ij} A_{ij}^{\mu}(t) |\dot{\mathbf{z}}_{\mu}(t)\rangle \otimes |ij\rangle\rangle, \qquad (29)$$

where $\mathbf{\Omega}$ is the inverse of the overlap matrix $S_{\mu\nu} = \langle \mathbf{z}_{\mu} | \mathbf{z}_{\nu} \rangle$. Based on Eqs. (26)-(29), the EOMs for the vibronic amplitudes can be derived as[128]

$$\sum_{\nu} S_{\mu\nu} \dot{A}_{ij}^{\nu} = -i \sum_{\nu, kl} \langle \mathbf{z}_{\mu} | \langle\langle ij | \hat{H}_{\text{eff}} | kl \rangle\rangle | \mathbf{z}_{\nu} \rangle A_{kl}^{\nu} - \sum_{\nu} \langle \mathbf{z}_{\mu} | \dot{\mathbf{z}}_{\nu} \rangle A_{ij}^{\nu}. \qquad (30)$$

For the effective Hamiltonian expressed in the normal ordering of $\{\hat{b}_k, \hat{b}_k^\dagger\}$, the matrix elements $\langle \mathbf{z}_\mu | \langle\langle ij | \hat{H}_{\text{eff}} | kl \rangle\rangle | \mathbf{z}_\nu \rangle$ can be easily calculated by Eq. (21). We adopt the ansatz of MCEv1 to get the EOMs of $\mathbf{z}_\mu$,[128]

$$\dot{\mathbf{z}}_\mu = -i \frac{1}{\langle \psi_\mu | \psi_\mu \rangle} \frac{\partial \langle \psi_\mu | \hat{H}_{\text{eff}} | \psi_\mu \rangle}{\partial \mathbf{z}_\mu^*}. \tag{31}$$

Eqs. (30) and (31) are the key equations for MCE-HEOM. Note that although we have used the MCEv1 ansatz, the proposed framework can also be directly combined with MCEv2.[131] As MCEv1 generally shows higher performance than MCEv2, we here only consider the conjecture of HEOM with MCEv1 and refer the method as MCE-HEOM.

The initial condition for MCE-HEOM can be directly obtained from that used in HEOM. At time zero, the zeroth-order ADO is set to $\hat{\rho}_S(0)$, while the remaining ADOs are set to 0 according to Eq. (12). The corresponding state vector can be expressed as

$$|\Psi(0)\rangle = \hat{\rho}_S(0) \otimes |0_1, 0_2, \cdots, 0_K\rangle, \tag{32}$$

where $0_k$ stands for the vacuum state of the $k$-th effective bath mode. Then the initial state vector can be projected to the coherent state basis as

$$\hat{\mathcal{P}}_T | \Psi(0) \rangle = \sum_{\mu=1}^{N_b} A^\mu(0) | \mathbf{z}_\mu(0) \rangle \otimes \hat{\rho}_S(0), \tag{33}$$

$$A^\mu(0) = \sum_\nu \Omega^{\mu\nu} \langle \mathbf{z}_\nu | 0_1, \cdots, 0_K \rangle. \tag{34}$$

By design, the initial values of $\mathbf{z}_\mu$ can be generated by sampling from a compressed Gaussian distribution[128]

$$P(\mathbf{z}_\mu) \propto e^{-\alpha_c |\mathbf{z}_\mu - \mathbf{z}_0|^2}, \tag{35}$$

where the bias is $\mathbf{z}_0 = 0$ in accordance with the vacuum state. As $\hat{\rho}_S(0)$ can be expanded in the Liouville space as $\hat{\rho}_S(0) = \sum_{ij} \rho_{ij}(0) |ij\rangle\rangle$, we may set $A_{ij}^\mu(0) = \rho_{ij}(0) A^\mu(0)$.

Applying Eq. (22), the system RDM can be given by projecting the state vector to the ground state of the effective modes as

$$\hat{\rho}_S(t) = \langle \mathbf{0} | \Psi(t) \rangle = \sum_{\mu=1}^{N_b} \sum_{ij} A_{ij}^\mu(t) \langle \mathbf{0} | \mathbf{z}_\mu(t) \rangle |ij\rangle\rangle = \sum_{\mu=1}^{N_b} \sum_{ij} A_{ij}^\mu(t) e^{-\frac{1}{2}\sum_k |z_\mu^{(k)}|^2} |ij\rangle\rangle. \quad (36)$$

The observables of the reduced system can be further calculated as

$$\langle \hat{O} \rangle = \text{Tr}_S \left[ \hat{O} \hat{\rho}_S(t) \right] = \langle \mathbb{I} | \hat{O} \sum_{\mu=1}^{N_b} \sum_{ij} A_{ij}^\mu(t) e^{-\frac{1}{2}\sum_k |z_\mu^{(k)}|^2} |ij\rangle\rangle, \quad (37)$$

where the identity vector is defined as[161]

$$|\mathbb{I}\rangle = \sum_i |ii\rangle\rangle \quad (38)$$

in order to perform the same operation as $\text{Tr}_S[\cdot]$.

For a quantum dynamics method, we may expect that the trace of the (reduced) density matrix should be preserved. For HEOM using the second quantization, the trace of the RDM is given by $\langle \mathbf{0}; \mathbb{I} | \Psi \rangle$ instead of $\langle \Psi | \Psi \rangle$, such that the trace conservation property should be reexamined. It can be easily proved that the trace of $\hat{\rho}_S$ is preserved when $|\Psi(t)\rangle$ is exactly evolved by Eq. (19).[122] However, tensor network based numerical solvers seem to conserve the trace of $\hat{\rho}_S$ only in the complete basis limit.[90,94] For MCE-HEOM, we here provide a proof in the appendix that when the system-bath interaction is absent, the EOMs of the RDM reduces to the Liouville equation subject to the system Hamiltonian $\hat{H}_S$, thus preserving the trace of $\hat{\rho}_S$. Thereby, our MCE-HEOM can reproduce the exact dynamics at least in one of the limiting cases.

## III. RESULTS AND DISCUSSION

The two-state spin-boson models[167] have been widely utilized to benchmark the performance of quantum dynamics methods. The system Hamiltonian reads

$$\hat{H}_S = \epsilon \hat{\sigma}_z + \Delta \hat{\sigma}_x, \tag{39}$$

where $\hat{\sigma}_z = |0\rangle\langle 0| - |1\rangle\langle 1|$ and $\hat{\sigma}_x = \Delta(|0\rangle\langle 1| + |1\rangle\langle 0|)$ are the Pauli matrices, $2\epsilon$ defines the energy bias between the two states, and $\Delta$ is the electronic coupling. In this paper, $\Delta$ is uniformly set to 1. The system-bath interaction term is written as

$$\hat{H}_{SB} = \hat{\sigma}_z \otimes \sum_j g_j \left( \hat{a}_j + \hat{a}_j^\dagger \right), \tag{40}$$

such that the dissipative system mode operator $\hat{Q}$ is taken to be $\hat{\sigma}_z$. For convenience, we here choose the Debye spectral density,

$$J(\omega) = \frac{\eta \omega_c \omega}{\omega^2 + \omega_c^2}, \tag{41}$$

as the corresponding exponents can be directly obtained from its poles. For other forms of spectral densities, frequency-domain or time-domain decomposition schemes can be used.[168–172] For the spin-boson model, Eq. (31) can be further simplified as

$$\dot{z}_\mu^{(k)} = -\gamma_k z_\mu^{(k)} - \mathrm{i} \frac{\sum_{ij} |A_{ij}^\mu|^2 \left[ d_k (-1)^i - d_{\bar{k}}^* (-1)^j \right]}{\sqrt{|d_k|} \sum_{ij} |A_{ij}^\mu|^2}, \tag{42}$$

where the first term $-\gamma_k z_\mu^{(k)}$ corresponds to the potential of $k$-th harmonic oscillator with a dissipative complex frequency for the $\mu$-th coherent state, and the remaining term is related to the system-bath interaction. If Eq. (42) is approximated as

$$\dot{z}_\mu^{(k)} = -\gamma_k z_\mu^{(k)}, \tag{43}$$

it can be regarded as the classical path approximation (CPA)[173] for complex-frequency cases. In this paper, we will show the results obtained with both Eqs. (42) and (43) to reveal the effect of coherent-state EOMs to the dynamics of RDM. To distinguish the two versions, MCE-HEOM with the CPA EOMs is denoted as MCE-HEOM-CPA. We also consider different numbers of $N_b$ to investigate the convergence of MCE-HEOM results. For MCE-HEOM-CPA, the results are uniformly calculated with $N_b = 20$. The initial state of the system is set on state $|0\rangle$, and the time-dependent RDM is utilized to benchmark the performance of different methods. The exact results are calculated by the mpsqd package using the 1-site TDVP method, which is denoted as MPS-HEOM.[174] The Matsubara decomposition is used for the four models presented in Figs. 1-4 with relatively high temperatures, while the Padé decomposition is used for the spin-boson model in Fig. 5 with a low temperature.[175,176] For comprehensive comparison, we also show the results obtained by MCEv1 with $N_b = 100$, which has a much larger $N_b$ than MCE-HEOM calculations. In all MCE-related calculations, the spectral densities of all the investigated spin-boson models are discretized into 100 modes.

As shown in Fig. 1, we first consider a representative symmetric spin-boson model proposed in ref. 174. Using the exact solutions by MPS-HEOM as references, MCEv1 obtains accurate population (i.e., $\rho_{00}(t)$) and imaginary part of coherence (i.e., Im$\rho_{01}(t)$) when 500 parallel realizations of the initial Boltzmann sampling are utilized to describe the finite temperature. In contrast, the real part of coherence (i.e., Re$\rho_{01}(t)$) by MCEv1 shows much larger deviations, implying that the coherence may require more initial samplings to get converged results or other hidden problems may exist in the MCEv1 method. Similarly, MCE-HEOM-CPA can give proper description of $\rho_{00}(t)$ and Im$\rho_{01}(t)$, but Re$\rho_{01}(t)$ shows much more evident deviations. In particular, the long-time Re$\rho_{01}(t)$ obtained by MCE-HEOM-CPA is completely wrong, which indicates that the good results of $\rho_{00}(t)$ and Im$\rho_{01}(t)$ might be a coincidence coming from error cancelation for the symmetric case. For the MCE-HEOM method with Eq. (42), we show the results with different numbers of coherent states as the basis set. It is apparent that there exist significant deviations in MCE-HEOM with $N_b = 1$, which can be regarded as a mean field approximation for the effective Hamiltonian. But it seems to behave differently compared to the mean field approximation for the conventional Hermitian Hamiltonian as the infinite temperature problem for long-time dynamics[177] does not appear in Fig. 1A. As we further increase $N_b$ to 10 and 20, it is encouraging that MCE-HEOM yields almost perfect dynamics for both population and coherence.

Compared to the conventional MCEv1 dynamics, MCE-HEOM does not need the initial Boltzmann sampling. In addition, the size of coherent state basis and the number of effective bath modes are significantly reduced, all resulting in a huge reduction of the computational cost. Considering that MCE-HEOM can reproduce the exact results with only 10 coherent states, the total number of parameters is only $(4+K) \times 10 = 90$, where $K = 5$ in this model. In comparison, as the MPS rank is $r = 20$ and the bond dimension is $d = 10$, the number of variational parameters for a single tensor in MPS-HEOM calculation is $d \times r^2 = 4000$,[174] which is already over 40 times larger than that used in MCE-HEOM. Although other tensor network state approaches may also reduce the number of variational parameters,[92] MCE-HEOM is free of the structure optimization of high-dimensional tensors and thus is much easier to implement.

In Fig. 2, we further study a more complex asymmetric model, whose parameters are taken from ref. 91. In this case, the traditional MCEv1 reasonably describes the population and coherence dynamics with 500 initial Boltzmann samplings, while some small oscillations and deviations are still observable. As a large energy bias is present,

MCE-HEOM-CPA yields an incorrect long-time limit for both $\rho_{00}(t)$ and $\text{Re}\rho_{01}(t)$, while $\text{Im}\rho_{01}(t)$ seems to be satisfactory, which also agrees with the discussions above. In contrast, MCE-HEOM still accurately reproduces both population and coherence dynamics for this asymmetric model with only 10 coherent states.

Considering the encouraging results of MCE-HEOM above, we further study two spin-boson models with strong non-Markovian effects taken from refs. 160 and 92. In these cases, the characteristic frequencies $\omega_c$ are relatively low, such that the bath responses slowly and presents a strong memory effect. It is important to note that the spin-boson model investigated in Fig. 3 has been considered as a tough benchmark system,[160,178,179] which requires a deep hierarchical tier (i.e., $L = 39$) for HEOM to get converged results.[178] As only one effective bath mode is necessary for this model, this problem can still be solved by the conventional HEOM. While MCEv1 yields incorrect long-time population and coherence dynamics with even 1000 initial Boltzmann samplings, it is a bit surprising that MCE-HEOM-CPA can nearly reproduce the exact results in this case. This is because the model has relatively classical bath modes ($\beta\Delta = 0.5$ and $\omega_c/\Delta = 0.25$) with a not very large system-bath interaction ($\eta/\Delta = 0.5$). Again, MCE-HEOM with 10 coherent states is enough to reproduce the exact solutions with high accuracy.

Fig. 4 shows the results for a more challenging model, which has a much lower characteristic frequency ($\omega_c/\Delta = 0.05$) and needs an even deeper hierarchical tier ($L = 50$) to get converged HEOM results.[92] Moreover, the bath is decomposed into 4 effective modes, eventually resulting in a huge number of ADOs for HEOM calculations. In this case, MCEv1 shows only small deviations for both $\rho_{00}(t)$ and Im$\rho_{01}(t)$ with 1000 initial Boltzmann samplings, but completely fails to give the correct long-time Re$\rho_{01}(t)$. As the system-bath interaction ($\eta/\Delta = 1$) is also larger than the model presented in Fig. 3, MCE-HEOM-CPA results in significant deviations for both population and coherence, which means that the feedback from the system to the bath is important in this case. In comparison, MCE-HEOM still yields accurate results with 10~20 coherent states.

In Fig. 5, we finally investigate another difficult case for HEOM, i.e., spin-boson models with low temperature. As the temperature ($\beta\Delta = 50$) is much lower than all cases studied above, a larger number of effective modes ($K = 13$) is required to reproduce the BCF even with the Padé spectrum decomposition method.[178,180] For MCEv1 calculations, 100 initial Boltzmann samplings are enough to get relatively accurate results, but the oscillation amplitudes are smaller than the exact references and Re$\rho_{01}(t)$ also shows some deviations at the long-time limit. In comparison, MCE-HEOM-CPA gives apparently too large oscillation amplitudes. This model also brings about larger challenges to MCE-HEOM, as the converged results with $N_b = 20$ and larger $N_b$ still show some deviations (although not large) with respect to the exact quantum references. As our MCE-HEOM is based on the MCEv1 ansatz, the EOMs of coherent states may not be perfect for the case with low-temperature and high characteristic frequency ($\beta\Delta = 50$ and $\omega_c/\Delta = 5$) mostly due to the nuclear tunneling effect.

## IV. CONCLUSIONS AND PERSPECTIVES

In summary, we have proposed a novel and robust MCE-HEOM method for the simulation of open quantum system dynamics, which has combined the key advantages of both MCE and HEOM. Based on the second quantization form of HEOM, the EOMs for MCE dynamics have been derived similar to those used for the traditional TDSE. As benchmarked in a series of representative and challenging spin-boson models, our MCE-HEOM has achieved accurate results very close to the exact solutions even when dealing with the problems requiring deep hierarchy tiers and having strong non-Markovian effects. Due to the high performance, MCE-HEOM can be regarded as a highly accurate and efficient solution to open quantum system dynamics and can be potentially utilized to investigate more complex systems in the future.[181–184]

Along the proposed framework in this study, some perspectives for future studies can be made. For instance, the pseudomode model can map the open quantum system to a core system composed of the system and pseudomodes, and the pseudomodes are coupled to a Markovian bath.[163,185–191] This approach uses the Lindblad equation[192–194] and might have more physical intuition than HEOM. The relation between the pseudomode model and HEOM has been confirmed by Li and Lv by introducing an ordered density operator (ODO).[122] Based on the definition of ODO, Su and coworkers developed an ordered moment approach and rederived HEOM, which bypasses the influence functional formalism.[195] The multiple Davydov ansatz has also been combined with the pseudomode model by Zeng and coworkers and the results are promising.[196] These studies imply that MCE can also be combined with the pseudomode model to give a more intuitive picture based on the Lindblad equation.

In MCE-HEOM, it is apparent that the system observables can be easily calculated based on the state vector in the composed Hilbert-Liouville space, and thus the high-order correlation functions of system operators can also be obtained, which can be used to study nonlinear spectroscopies.[17] For the bath operators, one may investigate whether MCE-HEOM can give related expectation values and correlation functions at least for the solvent modes (i.e., $\langle \hat{F}(t) \rangle$ and $\langle \hat{F}(t)\hat{F}(0) \rangle$, where $\hat{F}(t) = e^{i\hat{H}t}\hat{F}e^{-i\hat{H}t}$ should be distinguished from $\hat{F}_B(t)$). As pointed out in ref. 74 and based on the studies on DEOM, these quantities can be obtained through ADOs and the recursive relation.[74,79] It can be inferred from Eq. (22) that higher-order ADOs can also be obtained from the MCE ansatz. From a different perspective, based on generalized Langevin equations for the Gaussian bath, the system-bath entanglement theorem can be used to get general system-bath response functions from the local system response functions.[197–200] As the system response functions can be given directly by MCE-HEOM, the bath response functions can also be obtained through the system-bath entanglement theorem.

The semiclassical feature of the coherent states also indicates a new trajectory-based framework for the effective Hamiltonian given by HEOM. As Hamiltonian-based dynamics can be directly generated by the canonical relations, Ehrenfest-like mean filed equations can be derived easily from the effective Hamiltonian. We note that the second quantization form of HEOM has been recently combined with the classical mapping model with commutator variables (CMMcv) approach[201–207] by Zhang and coworkers.[208] It is also interesting to investigate whether other widely used nonadiabatic dynamics methods based on trajectories (e.g., trajectory surface hopping[209–214]) can be combined with HEOM. As each surface hopping trajectory should be evolved on an adiabatic potential energy surface, however, proper definitions of the eigenstates and nonadiabatic coupling in the case of Liouville space and non-Hermitian Hamiltonian[215] should be carefully considered.

We have only considered the bosonic bath in this study. As HEOM and its second quantization formalism also have the corresponding fermionic versions,[122,216–218] it is also possible to develop a fermionic MCE-HEOM. Nevertheless, the definition of coherent states for fermionic DoFs normally involves the Grassmann number,[146,147] which cannot be used directly for numerical programming.[219,220] Considering that there are some alternative choices for the construction of effective fermionic coherent states,[221–226] which is the most suitable scheme for an investigated problem still requires further investigation.

Finally, we note that MCE-HEOM also has some deficiencies. There is still no theoretical guarantee to ensure that MCE will give the exact results as the number of coherent states increases. In addition, the Liouville space-based framework prevents the applications of MCE-HEOM when the number of states in the system is too large. This is attributed to the fact that the number of elements in the density matrix scales quadratically compared to the linear scaling of the wave function. In this aspect, the stochastic Schrödinger equation formalism[160,227–233] in the Hilbert space might be a better choice for large systems. Note that the combination of hierarchy of pure states (HOPS) and MPS has already been proposed.[178,234–236] MCE can be similarly combined with HOPS. Moreover, a mixed deterministic-stochastic algorithm is expected to be superior than the pure deterministic strategy due to its capability for parallel computing.

The problem of decomposing BCFs can also be handled by the partial stochastic unraveling of the influence functional.[160,229] These related studies are currently underway in our group.

## ACKNOWLEDGMENTS

L.W. acknowledges support from the National Natural Science Foundation of China (Grant No. 22273082), Zhejiang Provincial Natural Science Foundation of China (Grant No. LZ25B030001), the Open Project Fund of National Facility for Translational Medicine (Shanghai) (Grant No. TMSK-2024-109), and the High-Performance Computing Center in Department of Chemistry, Zhejiang University.

## CONFLICT OF INTEREST

The authors have no conflicts to disclose.

## DATA AVAILABILITY STATEMENT

The data that support the findings of this study are available from the corresponding author upon reasonable request.

# APPENDIX: MCE-HEOM WITHOUT SYSTEM-BATH INTERACTION

It is apparent that an exact quantum dynamics method for open quantum systems should also be exact when describing the bare system dynamics without system-bath interaction. We here prove that the EOMs used in MCE-HEOM can exactly reduce to the Liouville equation for the system RDM in this limiting case.

When the system-bath interaction is zero, the effective Hamiltonian is given by

$$\hat{H}_{\text{eff}} = \mathcal{L}_{\text{S}} - i\sum_{k=1}^{K} \gamma_k \hat{b}_k^\dagger \hat{b}_k, \qquad (A1)$$

and the EOMs for $\mathbf{z}_\mu$ is simply Eq. (43) without the system feedback. Thereby, the matrix elements of $\langle \mathbf{z}_\mu | \dot{\mathbf{z}}_\nu \rangle$ can be simplified to

$$\langle \mathbf{z}_\mu | \dot{\mathbf{z}}_\nu \rangle = \sum_{k=1}^{K} \left( -\gamma_k z_\mu^{(k)*} z_\nu^{(k)} + \gamma_k \left| z_\nu^{(k)} \right|^2 \right) \langle \mathbf{z}_\mu | \mathbf{z}_\nu \rangle. \qquad (A45)$$

As the effective bath part in Eq. (A1) (i.e., $-i\sum_{k=1}^{K} \gamma_k \hat{b}_k^\dagger \hat{b}_k$) also produces similar terms, Eq. (30) can be further simplified as

$$\sum_\nu S_{\mu\nu} \dot{A}_{ij}^\nu = -i \sum_{\nu, mn} \langle ij | \mathcal{L}_{\text{S}} | mn \rangle S_{\mu\nu} A_{mn}^\nu - \sum_\nu \sum_{k=1}^{K} \gamma_k \left| z_\nu^{(k)} \right|^2 S_{\mu\nu} A_{ij}^\nu. \qquad (A46)$$

As the RDM in MCE-HEOM is given by Eq. (36), its time-derivative can be given based on Eq. (A46) as

$$\dot{\rho}_{ij} = \sum_\mu \left( \dot{A}_{ij}^\mu + \sum_{k=1}^{K} \gamma_k \left| z_\mu^{(k)} \right|^2 \right) e^{-\frac{1}{2}\sum_k \left| z_\mu^{(k)} \right|^2}. \qquad (A47)$$

Inverting the overlap matrix $S_{\mu\nu}$ in Eq. (A46) and then inserting it to Eq. (A47) give

$$\dot{\rho}_{ij} = -i\sum_{\mu, mn} \langle ij | \mathcal{L}_{\text{S}} | mn \rangle A_{mn}^\mu e^{-\frac{1}{2}\sum_k \left| z_\mu^{(k)} \right|^2} = -i\sum_{mn} \langle ij | \mathcal{L}_{\text{S}} | mn \rangle \rho_{mn}. \qquad (A48)$$

Here, Eq. (36) is used again. Eq. (A48) is exactly the Liouville equation written in the Liouville-space basis.

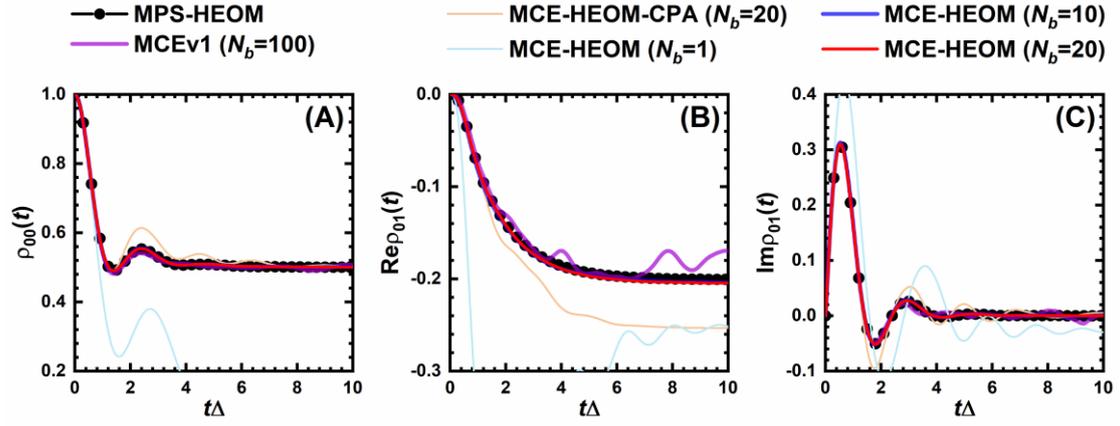

**Figure 1.** Time-dependent RDM calculated by different methods: (A) the population of state $|0\rangle$, (B) real and (C) imaginary parts of the coherence for the spin-boson model with $\epsilon/\Delta = 0$, $\eta/\Delta = 1$, $\beta\Delta = 0.5$, and $\omega_c/\Delta = 1$.

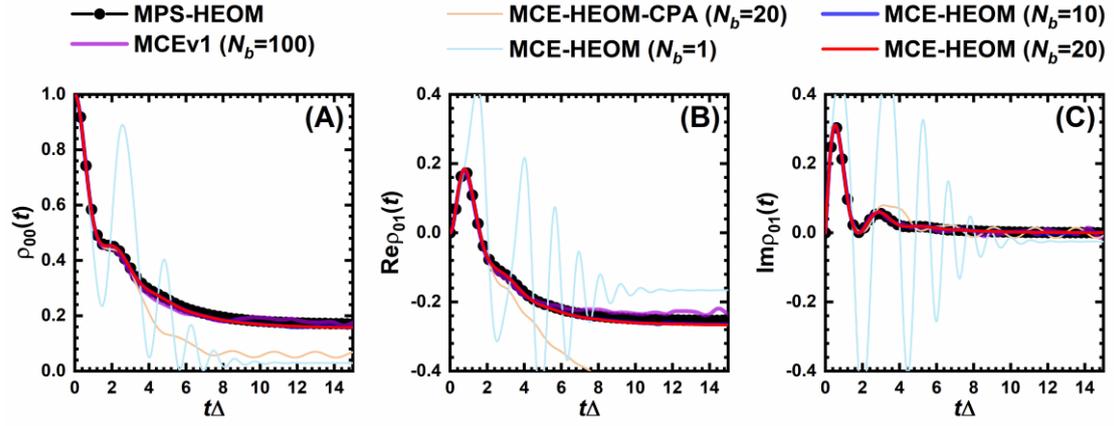

**Figure 2.** Time-dependent RDM calculated by different methods: (A) the population of state $|0\rangle$, (B) real and (C) imaginary parts of the coherence for the spin-boson model with $\epsilon/\Delta = 1$, $\eta/\Delta = 1$, $\beta\Delta = 1$, and $\omega_c/\Delta = 1$.

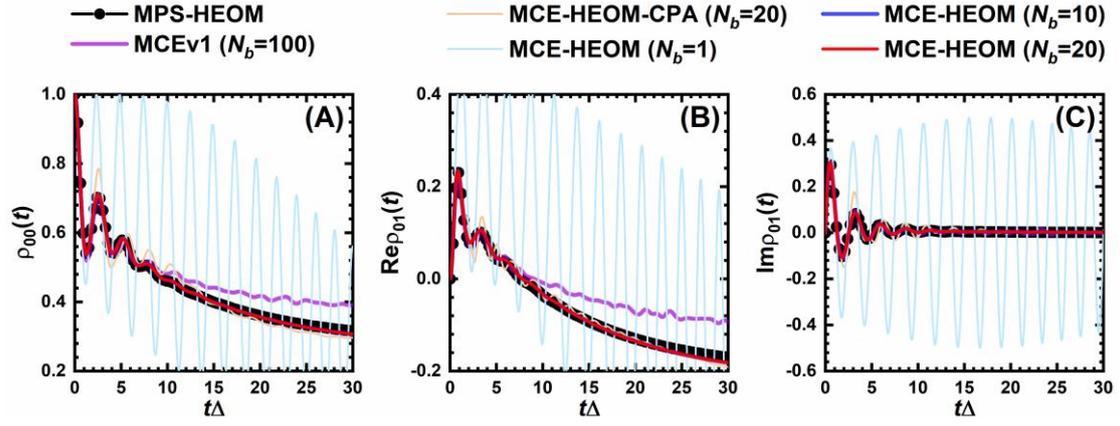

**Figure 3.** Time-dependent RDM calculated by different methods: (A) the population of state $|0\rangle$, (B) real and (C) imaginary parts of the coherence for the spin-boson model with $\epsilon/\Delta = 1$, $\eta/\Delta = 0.5$, $\beta\Delta = 0.5$, and $\omega_c/\Delta = 0.25$.

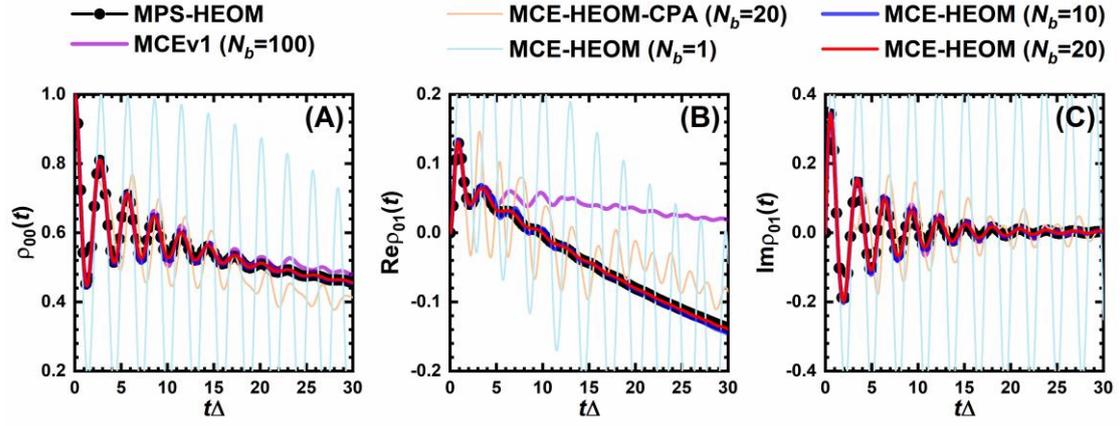

**Figure 4.** Time-dependent RDM calculated by different methods: (A) the population of state $|0\rangle$, (B) real and (C) imaginary parts of the coherence for the spin-boson model with $\epsilon/\Delta = 0.5$, $\eta/\Delta = 1$, $\beta\Delta = 0.96$, and $\omega_c/\Delta = 0.05$.

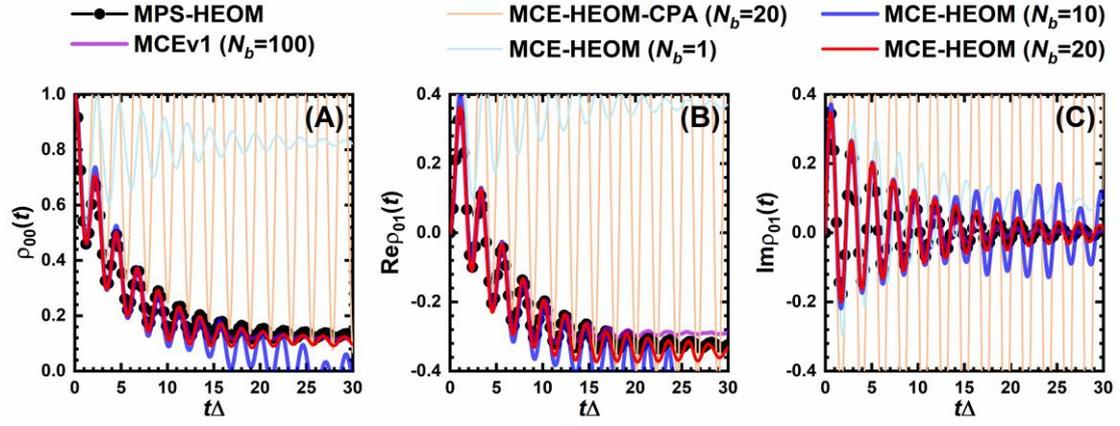

**Figure 5.** Time-dependent RDM calculated by different methods: (A) the population of state $|0\rangle$, (B) real and (C) imaginary parts of the coherence for the spin-boson model with $\epsilon/\Delta = 1$, $\eta/\Delta = 0.5$, $\beta\Delta = 50$, and $\omega_c/\Delta = 5$.